\newif\ifDraft\Draftfalse

\documentclass[10pt,nocopyrightspace]{sigplanconf}
\title{microAdapton}
\usepackage{alltt}
\usepackage{color}
\usepackage{graphicx}
\usepackage{natbib}
\usepackage{url}

\newcommand{\true}[1] {}
\newcommand{\false}[1] {}
\newcommand{\fixit}[1] {\ifDraft{\textcolor{red}{Fix: #1}}\fi}

\begin{document}
\bibliographystyle{plainnat}

\setlength{\pdfpageheight}{\paperheight}
\setlength{\pdfpagewidth}{\paperwidth}



\newcommand{\Matt}[1]{\ifDraft{{\color{blue}{Matt: #1}}}\fi}

\titlebanner{Scheme `16 --- September 18, 2016, Nara, Japan} 
\preprintfooter{Scheme `16}   

\title{miniAdapton}
\subtitle{A Minimal Implementation of Incremental Computation in Scheme}

\authorinfo{Dakota Fisher\and Matthew A. Hammer}
           {University of Colorado Boulder}
           {first.last@colorado.edu}
\authorinfo{William Byrd\and Matthew Might}
           {University of Utah}
           {Will.Byrd@utah.edu\and might@cs.utah.edu}

\maketitle

\begin{abstract}
  We describe a complete Scheme implementation of \emph{miniAdapton},
  which implements the core functionality of the Adapton system for
  incremental computation (also known as \emph{self-adjusting
    computation}).
  Like Adapton, miniAdapton allows programmers to safely combine
  mutation and memoization.
  miniAdapton is built on top of an even simpler system,
  \emph{microAdapton}.
  Both miniAdapton and microAdapton are designed to be easy to
  understand, extend, and port to host languages other than Scheme.
  We also present \emph{adapton variables}, a new interface in Adapton
  for variables intended to represent expressions.
\end{abstract}




\section{Introduction}
\fixit{We need to check our typesetting and make sure we do not overflow with our code.}
\fixit{Also, I need someone check the conference information and copyright information, I still haven't put the date in full}

Memoization~\citep{michie:1968} is a simple yet powerful optimization technique, avoiding redundancy in computation to save considerable amounts of time.
When used properly, memoization can achieve asymptotic speedup of many algorithms.  Amazingly, in some cases memoization can even transform an exponential-time program into a linear-time program~\citep{clrs}.
Unfortunately, although memoization is an extremely powerful technique, it suffers from a serious limitation: memoization does not work in the presence of mutation.

Consider \texttt{max-tree} and \texttt{max-tree-path}, two memoized functions (see Appendix~\ref{sec:memo} for a complete definition of \texttt{define-memo}).
Function~\texttt{max-tree} finds the maximum number in a tree made of pairs and numbers and \texttt{max-tree-path} finds the path from the root of the tree to the maximum number.

\begin{alltt}
(define-memo (max-tree t)
  (cond
    ((pair? t)
     (max (max-tree (car t))
       (max-tree (cdr t))))
    (else t)))

(define-memo (max-tree-path t)
  (cond
    ((pair? t)
     (if (> (max-tree (car t))
            (max-tree (cdr t)))
         (cons 'left (max-tree-path (car t)))
         (cons 'right (max-tree-path (cdr t)))))
    (else '())))
\end{alltt}

\noindent
Observe that following the path returned by \texttt{max-tree-path} should yield the value returned by \texttt{max-tree}.

Suppose that we also have a tree, \texttt{some-tree}:
\begin{alltt}
(define some-tree '((1 . 2) . (3 . 4)))
\end{alltt}

\noindent
For clarity, here is \texttt{some-tree} explicitly represented as a tree:\\

\begin{center}
\vspace{-3.5mm}
\includegraphics[width=0.25\columnwidth]{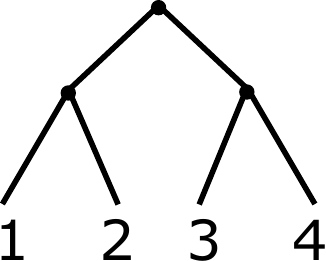}
\vspace{-1mm}
\end{center}

Now let's consider a user session.  First we ask for the maximum value of any leaf node in \texttt{some-tree}:
\begin{alltt}
> (max-tree some-tree)
4
\end{alltt}
This is what we expect, since the maximum number in the tree is clearly 4.

Next we modify \texttt{some-tree}, replacing its entire right-hand branch with the number 5.
\begin{alltt}
> (set-cdr! some-tree 5)
> some-tree
((1 . 2) . 5)
\end{alltt}

\noindent
And here is the updated explicit tree representation of \texttt{some-tree}:

\begin{center}
\vspace{-3.5mm}
\includegraphics[width=0.20\columnwidth]{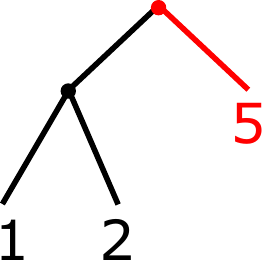}
\vspace{-1mm}
\end{center}

Once more we ask for the maximum number in the tree.

\begin{alltt}
> (max-tree some-tree)
4
\end{alltt}
This answer is no longer correct, since now the maximum number in the tree is 5, not 4.  In fact, 4 isn't even in the tree anymore!
As soon as the user performs a single mutation in the tree, our statements about the supposed behavior of the program are promptly violated.
\emph{Memoization cannot handle mutation.}

We can use \texttt{max-tree-path}, which determines the path to the maximum value, to figure out what went wrong.
\begin{alltt}
> (max-tree-path some-tree)
(right)
> (max-tree (cdr some-tree))
5
\end{alltt}
Following \texttt{max-tree-path} to the cdr of the tree yields a subtree containing a larger number than the entire tree!

In our implementation of memoization, shown in Appendix~\ref{sec:memo}, the problem is even worse, since the user has mutated the lookup key.
Now \emph{every} instance of \mbox{\texttt{((1 . 2) . 5)}} irreparably has the maximum number 4 somewhere in the tree according to the function, which is apparently in the path (right), where a 5 resides.


In order to make memoization work in the presence of mutation, we must keep additional information about the computation while also keeping track of the mutations which have occured. %
\emph{Adapton}~\citep{Hammer:2014:ACD:PLDI} is a system that provides these capabilities, combining the flexibility of mutation with the potentially asymptotic speedup of memoization.
Adapton is a form of \emph{incremental computation} (sometimes called \emph{self-adjusting computation}), an umbrella term for techniques which take changing inputs and save time by reusing previous results to compute new results for mutated inputs~\citep{DBLP:journals/entcs/AcarBBHT06}.

Although Adapton reconciles memoization and mutation, the complete code for an implementation of Adapton has not appeared in the literature.  In addition, Adapton implementations have focused on performance rather than on understandability, portability, or hackability.  Inspired by microKanren~\citep{Hemann:SW:2013}, which presented a tiny, easily understandable implementation of the core of the miniKanren programming language~\citep{Byrd:2006fk}, we have created \emph{microAdapton}, a tiny, easily understandable implementation of the core of Adapton.  Also in the spirit of microKanren and miniKanren, we build a higher-level interface, \emph{miniAdapton}, on top of the microAdapton core.

Our paper provides the first complete implementation of a version of Adapton in the literature.  More specifically, our paper presents two small implementations of Adapton:
\begin{itemize}
\item microAdapton (section~\ref{sec:micro}): a minimal substrate on which to build the miniAdapton system, providing only the barest interfaces to aid portability and readability.
Separating miniAdapton into microAdapton and miniAdapton allows us to provide a simple and portable core and layer user functionality on top of it.
Inspired by microKanren, microAdapton contains zero macros and fits in under 50 lines of source code.
It avoids building Adapton's Demanded Computation Graph, relegating it to be done by miniAdapton, or manually if used directly, making for a more flexible yet less immediately useful system.
\item miniAdapton (section~\ref{sec:mini}): a complete implementation of adapton, intended to provide a full Adapton system to the user, including:
\begin{itemize}
\item an interface to adaptons which automatically maintains Adapton's Demanded Computation Graph (section~\ref{subsec:force});
\item functions and macros for function memoization (section~\ref{subsec:amemo});
\item a convenience macro for constructing elements of adapton thunks representing expressions similar to delay (section~\ref{subsec:adapt});
\item adapton variables: a new interface in Adapton for variables intended to represent expressions (section~\ref{subsec:avar}).
\end{itemize}

\end{itemize}

The code for these implementations is publicly available here:
\url{https://github.com/fisherdj/miniAdapton}


\section{Overview}
Adapton, like memoization, stores the results of computation so that it can avoid redundant computation.
\true{DJF this claim is simply true, and is backed up by miniAdapton.}
Memoization, however, stores only the results, which means that even after mutation, the computation cannot be restarted, and even if it could, it might not be possible to figure out when it needs to be restarted.
\true{DJF true claim, backed up by our implementation, assuming we properly reflect memoization}
Instead of storing just the results of computations, Adapton stores their results, how it got those results and a graph of the dependencies between those computations.
Then, whenever a value is mutated through the Adapton interface, the computations depending on that value are marked dirty to indicate that their computation must be restarted.
\false{DJF previously erroneous: I have qualified that a value must be mutated through the adapton interface}
This lets Adapton keep many of the benefits of memoization while still permitting mutation.
\fixit{DJF: I'm tempted to reverse the order of presentation of these three subsections. This way, I can present the DCG upfront, describe super/subcomputation with visual aid of the DCG, and then use this to motivate from-scratch consistency, where mutable elements are the ``leaves'' of the DCG}
\Matt{The proposed arrangement sounds good; in particular, it seems best to begin by showing the DCG for the example. In the context of this example, we can explain the general structure of the DCG.}
\subsection{What is Adapton?}
Adapton refers to a system for self-adjusting computation by keeping track of computation at runtime.
An adapton thunk (or athunk) keeps track of any subcomputations it needs to determine its value and any supercomputations that need its value.
The athunks maintain the property of \emph{From-scratch Consistency}, meaning that after modifying referenced mutable elements through Adapton, the result of forcing these thunks is the same as if one had computed them from scratch.
\true{DJF true claim}
However, because athunks keep track of computation. they can avoid redundant re-computation, yielding the potential speedup of memoization while retaining the capability of mutation (see~\citep{Hammer:2014:ACD:PLDI} for performance discussion).
\false{DJF mostly true claim, an efficient implementation of memoization will always be at least as good as an efficient implementation of adapton, but we still retain asymptotic speedup}

Here athunks are mutable promises: they store a computation which may reference mutable objects in memory (adapton-refs) and cache the result of the computation but when forced will always return the same value as if computed from scratch.
Other behaviors may be possible, but this is the primary use case.

\subsection{Supercomputation and Subcomputation}
If $a$ is some computation (an athunk), then $a$'s supercomputations are computations that depend on $a$, and $a$'s subcomputations are computations that $a$ depends on.
If $b$ is $a$'s subcomputation, then $a$ is $b$'s supercomputation.

\subsection{Demanded Computation Graph}
The demanded computation graph or DCG is a graph representing the dependencies between computations and their supercomputation/subcomputation relation.
Each node of the graph is a computation and each edge is a supercomputation/subcomputation relation.
This implementation provided uses inefficient implementations of sets and memoization, which may hinder its performance (see Sections \ref{sec:sets} and \ref{sec:memo} for discussion on speeding it up).

\fixit{I explained this graph, and need the explanation reviewed.}
\Matt{I reviewed the graph, it looks right to me; I did a pass over the text below, which looks correct; though, maybe this description should be more detailed, and go through the steps of updating the graph in more step-by-step detail.}
The figure below is a representation of the DCG for the computation \mbox{\texttt{(max-tree some-tree)}} when \texttt{some-tree} is modified from \mbox{\texttt{((1 . 2) . (3 . 4))}} to \mbox{\texttt{((1 . 2) . 5)}}.
Nodes represent athunks and edges represent supercomputation/subcomputation relationships: An edge exists from node~$n$ to node~$m$ when $n$ depends on $m$ (i.e., this edge represents that $n$ is the supercomputation and $m$ is one of its subcomputations).
Black nodes and edges represent portions of the original graph, red nodes and edges represent portions of the original graph which are dirtied or which propagate dirtiness, and gray nodes and edges will be created when the adapton is forced again later.
Modifying {\tt some-tree} dirties the root node;  when forced again it will create new thunks for \mbox{\texttt{(max-tree '((1 . 2) . 5))}} and \mbox{\texttt{(max-tree 5)}} to complete the computation.
Although the root node still points to \mbox{\texttt{(max-tree '((1 . 2) . (3 . 4)))}}, the edge between the two is removed before the thunk is recomputed, which regenerates the edge anew.

\includegraphics[width=\columnwidth]{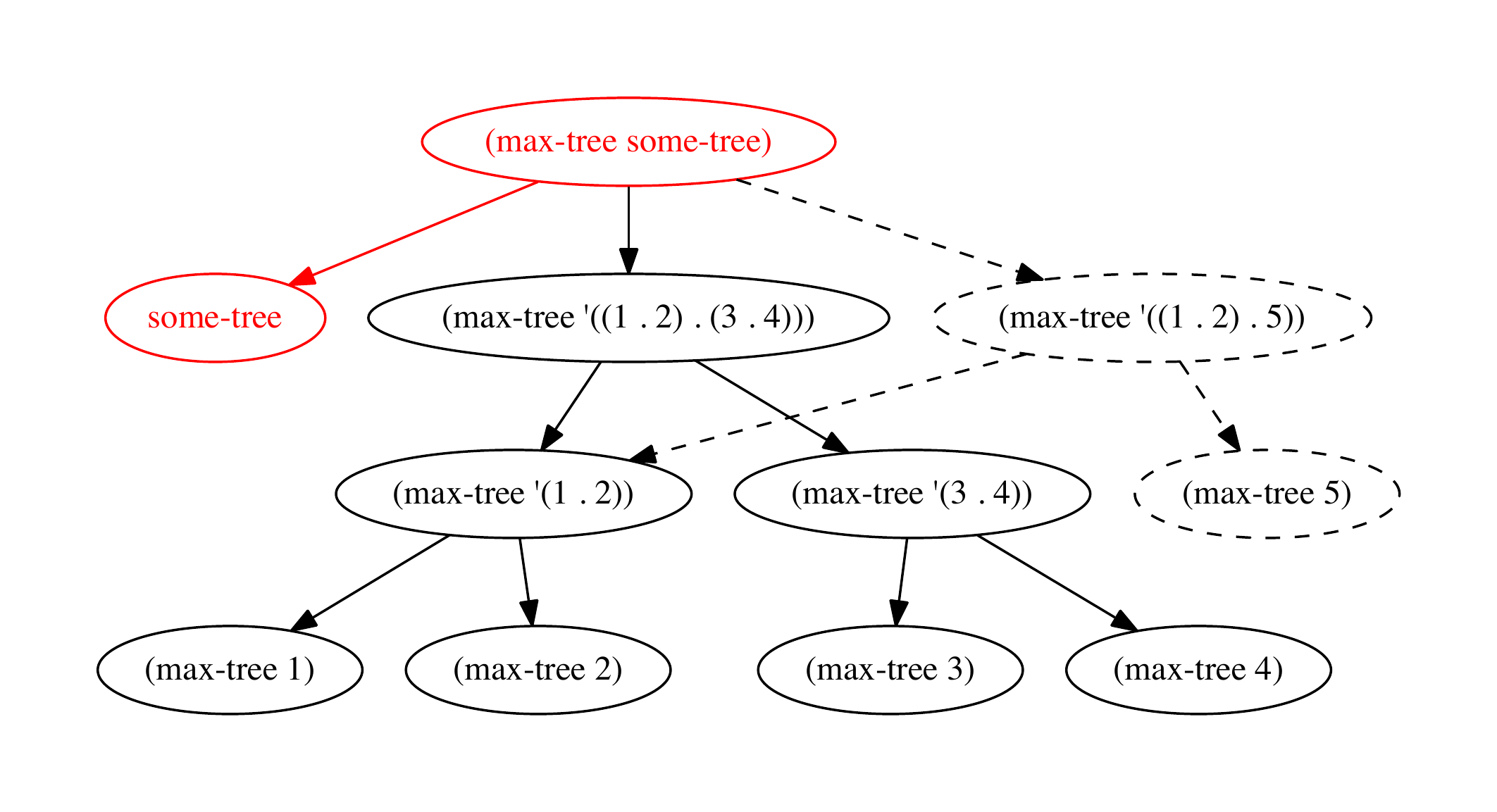}

\Matt{Say more? Walk through the update step by step?}

\section{microAdapton Implementation}
\label{sec:micro}
The implementation of microAdapton operates primarily on the \texttt{adapton} data structure, the datatype is defined as:
\begin{alltt}
(define-record-type
  (adapton adapton-cons adapton?)
  (fields
   thunk
   (mutable result)
   (mutable sub)
   (mutable super)
   (mutable clean?)))
\end{alltt}
The fields are:
\begin{description}
\item[thunk] the computation to cache
\item[result] cached result of the computation
\item[sub] set of subcomputations
\item[super] set of supercomputations
\item[clean?] whether or not the cached result is valid
\end{description}

Function~\texttt{make-athunk} constructs an athunk representing a thunk yet to be computed.
\begin{alltt}
(define (make-athunk thunk)
  (adapton-cons thunk
                'empty
                empty-set
                empty-set
                #f))
\end{alltt}
The first parameter of the constructor is the only parameter of the \texttt{make-athunk} function.
The result is arbitrarily set to the symbol \texttt{empty}.
The sub and super sets are both empty, since this athunk is not yet placed in the DCG.
Finally, the athunk is not clean until after the thunk is computed, so cleanliness is false.

Function~\texttt{adapton-add-dcg-edge!} adds edges of the DCG, and \texttt{adapton-del-dcg-edge!} removes edges of the DCG, from their parameters \texttt{a-super} and \texttt{a-sub}.
\begin{alltt}
(define (adapton-add-dcg-edge! a-super a-sub)
  (adapton-sub-set!
   a-super
   (set-cons a-sub (adapton-sub a-super)))
  (adapton-super-set!
   a-sub
   (set-cons a-super (adapton-super a-sub))))

(define (adapton-del-dcg-edge! a-super a-sub)
  (adapton-sub-set!
   a-super
   (set-rem a-sub (adapton-sub a-super)))
  (adapton-super-set!
   a-sub
   (set-rem a-super (adapton-super a-sub))))
\end{alltt}
Adding the DCG edge means connecting the athunks together by adding to their respective sub and super sets, while removing the DCG edge means disconnecting the athunks by removing their sub and super sets.

Function~\texttt{adapton-compute} computes an athunk and performs maintenance, returning the athunk's value, and keeping from-scratch consistency if the DCG is correctly maintained:
\begin{alltt}
(define (adapton-compute a)
  (if (adapton-clean? a)
      (adapton-result a)
      (begin
        (set-for-each
          (lambda (x)
            (adapton-del-dcg-edge! a x))
          (adapton-sub a))
        (adapton-clean?-set! a #t)
        (adapton-result-set! a
          ((adapton-thunk a)))
        (adapton-compute a))))
\end{alltt}
If the result is already available and valid, signified by the athunk being clean, then that value is returned.
Otherwise, either because the result is not available or because the current result is invalid, additional maintanence must be performed.
First, since computation is starting or restarting, all subcomputation relations are invalid and the subcomputation set must be removed.
Second, the athunk must be marked as clean.
Third, the result of the computation must be computed and stored in the athunk.
After this maintanence, we restart computing the athunk, which may lead to restarting the computation if the athunk has been marked dirty during the computation.
This maintains from-scratch consistency in certain edge cases where mutation occurs within computation.
\false{DJF: this is true, but little defended}
The order of maintanence operations is important: computing the result must be done after removing edges to subcomputations and after marking the athunk as dirty.
Function~\texttt{adapton-compute} is very sensitive to errors in the DCG.
If an edge is added to a dirty athunk as a subcomputation, changes will not propagate until that athunk is clean.
\fixit{Maybe I should have talked more about this in the overview}
This is a direct consequence of Adapton's lazy change propagation strategy: changes are indicated in the DCG, but the changes themselves are not computed until forcing the corresponding athunks.
Avoiding marking changes more than once prevents redundant repeated traversal of large DCGs.

Function~\texttt{adapton-dirty!} implements this change propagation strategy.
It marks an athunk dirty along with all supercomputations, under the assumption that the DCG is correctly maintained so that any dirty subcomputations are not subcomputations of clean nodes.
\begin{alltt}
(define (adapton-dirty! a)
  (when (adapton-clean? a)
        (adapton-clean?-set! a #f)
        (set-for-each adapton-dirty!
                      (adapton-super a))))
\end{alltt}
Calling \texttt{adapton-dirty!} on the immediate supercomputations of an athunk is sufficient for change propagation.
Since {\tt adapton-dirty!} only recurses for an athunk marked clean, calling {\tt adapton-dirty!} twice or more on any one athunk without computing it (and thus, cleaning it) between calls traverses the graph only once.
\true{DJF: this is true}

Finally, we need to be able to create mutable cells.
Function~\texttt{adapton-ref} takes a value creates a new ref cell with that value without needing to define an additional structure.
\begin{alltt}
(define (adapton-ref val)
  (letrec ((a (adapton-cons
               (lambda () (adapton-result a))
               val
               empty-set
               empty-set
               #t)))
    a))
\end{alltt}
Because it references itself, the ref can be set just by modifying the value in its cell.
Since this invalidates supercomputations, one must also call \texttt{adapton-dirty!} to guarantee from-scratch consistency.
\true{This is true, the wording is very precise here. It is necessary to provide the guarantee, it may incidentally be true. The only caveat is that it may be valid when a ref is set to its own value.}
Function~\texttt{adapton-ref-set!} sets the value of the ref cell in its first parameter to the second parameter.
\begin{alltt}
(define (adapton-ref-set! a val)
  (adapton-result-set! a val)
  (adapton-dirty! a))
\end{alltt}
Function~\texttt{adapton-ref-set!} is the sole method of mutation provided.


The microAdapton interface exposes these functions: \texttt{adapton-compute}, \texttt{adapton-ref}, \texttt{adapton-ref-set!}, \texttt{adapton?}, \texttt{make-athunk}, \texttt{adapton-add-dcg-edge!}, and \texttt{adapton-del-dcg-edge!}.
All uses and users of microAdapton, including the higher-level miniAdapton interface, must use this interface; modifying the DCG directly, without using the interface functions, may result in unspecified or erroneous behavior.
\fixit{I need to run this code}
\Matt{Have you done this yet?}
\fixit{Done, fixed errors in this session. I need to do a pass to make sure every example has a file and runs correctly.}
\begin{alltt}
(define r1 (adapton-ref 8))
(define r2 (adapton-ref 10))
(define a
  (make-athunk
    (lambda ()
     (adapton-add-dcg-edge! a r1)
     (adapton-add-dcg-edge! a r2)
     (- (adapton-compute r1)
        (adapton-compute r2)))))

> (adapton-compute a)
-2
> (adapton-ref-set! r1 2)
> (adapton-compute a)
-8
\end{alltt}
\fixit{
We need explanation of what's happening in the lambda.
We must add the DCG edges in the lambda itself because the edges are cleaned prior computation.
When we set r1, we dirty r1, which then dirties a.
When we compute a the second time, we see that it is dirty and restart computation.
}

\section{miniAdapton: A Higher-level Interface}
\fixit{I need some re-reading on this. Closer inspection revealed that I only mostly finished this section instead of entirely finishing it.}
\label{sec:mini}
\true{DJF: true, it is a higher-level interface}
A user of microAdapton has all the tools they need to get the benefits of from-scratch consistency and memoization.
The user must, however, build the DCG manually within computations.
microAdapton provides only the low-level operations adding and removing edges for correct construction and use of DCGs.
\false{DJF: this is true, but a stronger claim can be made. The ONLY way to build the DCG is to use these operations}
This process is tedious, mechanical, potentially error-prone and should be avoided where possible.
Truly memoizing functions using athunks requires significant additional work: one has to use \texttt{letrec} to grab the thunk to add it, and must reference the thunk inside the other function.
Thankfully, some extra interfaces can be made readily available to the user which avoids these difficulties.

\subsection{Essential Interface}
\label{subsec:force}

The essential interface to miniAdapton is \texttt{adapton-force}, a function that marks athunks as subcomputations if their computation occurs during the computation of a different athunk.
Although using \texttt{adapton-compute} directly cannot provide this, we can instead define \texttt{adapton-force} and use it exclusively in place of \texttt{adapton-compute}, as shown in this user session:

\begin{alltt}
(define r (adapton-ref 5))
(define a
  (make-athunk
    (lambda ()
     (+ (adapton-force r) 3))))

> (adapton-force a)
8
> (adapton-ref-set! r 2)
> (adapton-force a)
5
\end{alltt}

\true{true, adapton-force has the last instance of adapton-compute}

Here is the definition of \texttt{adapton-force}:

\begin{alltt}
(define adapton-force
  (let ((currently-adapting #f))
    (lambda (a)
      (let ((prev-adapting
             currently-adapting))
        (set! currently-adapting a)
        (let ((result (adapton-compute a)))
          (set! currently-adapting
                prev-adapting)
          (when currently-adapting
                (adapton-add-dcg-edge!
                 currently-adapting
                 a))
          result)))))
\end{alltt}
Function~\texttt{adapton-force} keeps track of any Adapton computation that we are immediately in and, when called with a single athunk argument, computes the result of its argument and places the athunk in the DCG.

\subsection{The adapt Form}
\label{subsec:adapt}
\fixit{Potentially switch this and previous section}
A useful macro styled after delay takes expressions and turns them into athunks.
Translating our example for adapton-force to use adapt:
\begin{alltt}
(define r (adapton-ref 5))
(define a (adapt (+ (adapton-force r) 3))))
\end{alltt}

The \texttt{adapt} macro is straightforward, it takes an expression, wraps it in a thunk and then packages it with \texttt{make-athunk}:

\begin{alltt}
(define-syntax adapt
  (syntax-rules ()
    ((_ expr)
     (make-athunk (lambda () expr)))))
\end{alltt}

\subsection{Memoization}
\label{subsec:amemo}
Function \texttt{adapton-force}, paired with a memoization implementation (Appendix~\ref{sec:memo}), allows us to define two memoization procedures for Adapton.
Here are \texttt{max-tree} (find maximum number in tree) and \texttt{max-tree-path} from the introduction (find path to maximum number in tree) translated into a variant utilizing Adapton:
\true{DJF: true}
\begin{alltt}
(define-amemo (max-tree t)
  (cond 
   ((adapton? t)
    (max-tree (adapton-force t)))
   ((pair? t)
    (max (max-tree (car t))
         (max-tree (cdr t))))
   (else t)))

(define-amemo (max-tree-path t)
  (cond
    ((adapton? t)
     (max-tree-path (adapton-force t)))
    ((pair? t)
     (if (> (max-tree (car t))
            (max-tree (cdr t)))
         (cons 'left
               (max-tree-path (car t)))
         (cons 'right
               (max-tree-path (cdr t)))))
    (else '())))
\end{alltt}

\true{reworded, the interface was enough of miniAdapton, but memoization is also necessary, Scheme provides everything else we need}

The following functions are convienences for defining memoized functions:

\begin{alltt}
(define (adapton-memoize-l f)
  (memoize (lambda x (adapt (apply f x)))))

(define (adapton-memoize f)
  (let ((f* (adapton-memoize-l f)))
    (lambda x (adapton-force (apply f* x)))))
\end{alltt}

The first function produces memoized functions returning athunks, where the ``l'' suffix means ``lazy.''
The second produces functions which are both memoized and from-scratch consistent within the adapton system.
Function~\texttt{adapton-memoize-l} operates by memoizing a version of its function argument returning athunks.
A standard implementation of memoization in conjunction with the \texttt{adapt} form is sufficient to perform this task.
Function~\texttt{adapton-memoize} operates by taking the result of \texttt{adapton-memoize-l} and making a new function equivalent except that it returns the result of forcing the athunk instead of the athunk itself.
\true{DJF: All claims in this paragraph are true}

A small battery of convenience macros for memoization are also useful to have.
We have two macros for constructing procedures with the syntax of \texttt{lambda}:
\begin{alltt}
(define-syntax lambda-amemo-l
  (syntax-rules ()
    ((_ (args ...) body ...)
     (let ((f* (adapton-memoize-l
                (lambda (args ...)
                  body ...))))
       (lambda (args ...) (f* args ...))))))

(define-syntax lambda-amemo
  (syntax-rules ()
    ((_ (args ...) body ...)
     (let ((f* (adapton-memoize
                (lambda (args ...)
                  body ...))))
       (lambda (args ...) (f* args ...))))))
\end{alltt}
and two for defining procedures with the syntax of \texttt{define}:
\begin{alltt}
(define-syntax define-amemo-l
  (syntax-rules ()
    ((_ (f args ...) body ...)
     (define f (lambda-amemo-l (args ...)
                 body ...)))))

(define-syntax define-amemo
  (syntax-rules ()
    ((_ (f args ...) body ...)
     (define f (lambda-amemo (args ...)
                 body ...)))))
\end{alltt}

\subsection{Adapton Variables}
Function~\texttt{adapton-ref-set!} can be a somewhat unwieldy interface, one can only set references equal to values, when we might want to set those references to expressions instead.
Consider these definitions:
\begin{alltt}
(define r1 (adapton-ref 2))
(define r2
  (adapton-ref (+ (adapton-force r1) 4)))
(define a
  (adapt (+ (adapton-force r1)
            (adapton-force r2))))
\end{alltt}
Notice that we force \texttt{r1} to get the value of \texttt{r2}.
Upfront, we expect that \texttt{r1} has 2 stored in it and \texttt{r2} has 6 stored in it.
Now consider this user session:
\begin{alltt}
> (adapton-force a)
8
> (adapton-ref-set! r1 10)
> (adapton-force a))
16
\end{alltt}
The first answer is not surprising, it corresponds to adding 2, 2 and 4 (8).
After we set and force a, one might desire to get the result of adding 10, 10 and 4 (24), but instead we get the result of adding 10, 2 and 4 (16).
This happens because \texttt{r2} is a ref and retains its value instead of updating to match the value of its expression.
\label{subsec:avar}
The final useful interfaces provided are the ``avar'' macros and functions, short for ``adapton variable.''
Adapton variables provide the behavior of acting as expressions rather than values:
\begin{alltt}
(define-syntax define-avar
  (syntax-rules ()
    ((_ name expr)
     (define name
             (adapton-ref (adapt expr))))))

(define (avar-get v)
  (adapton-force (adapton-force v)))

(define-syntax avar-set!
  (syntax-rules ()
    ((_ v expr)
     (adapton-ref-set! v (adapt expr)))))
\end{alltt}
An avar is a variable representing an expression which can be changed and will remain from-scratch consistent with other changes.
It is made of an adapton-ref that itself contains an athunk for the desired expression.
Macro~\texttt{define-avar} defines an avar and assigns it an expression.
Function~\texttt{avar-get} gets the value resulting from evaluating the avar's expression, forcing the adapton ref to get the thunk and then forcing that to actually obtain the value from the thunk.
Macro~\texttt{avar-set!} sets the expression of an avar.

\fixit{I just added this example}
Here is our example again, translated to use avars:
\begin{alltt}
(define-avar v1 2)
(define-avar v2 (+ (avar-get v1) 4))
(define-avar b
  (+ (avar-get v1) (avar-get v2)))

> (avar-get b)
8
> (avar-set! v1 10)
> (avar-get b))
24
\end{alltt}
This code is much easier to reason about, since we can get the correct value by only looking at the expressions.


\subsection{Putting it All Together: Extended Example}
Now that we have all of miniAdapton available to us, we can translate and extend the example of our introduction.
Before we do so, let's define a new function to help us inspect trees containing our new data structures:
\begin{alltt}
(define (remove-adapton v)
  (cond
    ((pair? v)
     (cons (remove-adapton (car v))
       (remove-adapton (cdr v))))
    ((adapton? v)
     (remove-adapton (adapton-force v)))
    (else v)))
\end{alltt}

This recursively takes any athunks in our structures and replaces them with their values.
Here are some avar definitions in addition to our original \texttt{some-tree}:
\begin{alltt}
(define-avar lucky 7)
(define-avar t1 (cons 1 2))
(define-avar t2 (cons 3 4))
(define-avar some-tree
  (cons (avar-get t1) (avar-get t2)))
\end{alltt}
Now our user session from the start:
\begin{alltt}
> (avar-get some-tree)
((1 . 2) 3 . 4)
> (max-tree some-tree)
4
> (max-tree-path some-tree)
(right right)
\end{alltt}
So far, so good, but this is where the original broke last time.
We do the same thing, setting the cdr of \texttt{some-tree} to 5:
\begin{alltt}
> (avar-set! t2 5)
> (avar-get some-tree)
((1 . 2) . 5)
> (max-tree some-tree)
5
> (max-tree-path some-tree)
(right)
> (max-tree (cdr (avar-get some-tree)))
5
> (max-tree-path (cdr (avar-get some-tree)))
()
\end{alltt}
The memoized functions are now reporting the correct results in spite of the mutation.
In fact, we have even more flexibility than our introduction example might suggest.
We can set our subtrees to expressions instead of values:
\begin{alltt}
> (avar-set! t2
    (cons 20 (* 3 (avar-get lucky))))
> (avar-get some-tree)
((1 . 2) 20 . 21)
> (max-tree some-tree)
21
> (max-tree-path some-tree)
(right right)
\end{alltt}
This is all as usual, \texttt{t2} has the value \mbox{\texttt{(20 . 21)}}.
\begin{alltt}
> (avar-set! lucky 3)
> (avar-get some-tree)
((1 . 2) 20 . 9)
> (max-tree some-tree)
20
> (max-tree-path some-tree)
(right left)
\end{alltt}

Even in the presence of more complicated dependencies of computation, from-scratch consistency is maintained.
In addition, the code is made easier to reason about, all the properties that were lacking in our introductory example are once again present.

\section{Related Work}

Researchers have provided many language-based
approaches to incremental
computation~\citep{DBLP:journals/entcs/AcarBBHT06,HammerAc08,AcarLW09,NominalAdapton2015}.
In particular, researchers have shown that for certain algorithms,
inputs, and classes of input changes, IC delivers large, even
\emph{asymptotic} speed-ups over full
reevaluation~\citep{AcarIhMeSu07, AcarBlTaTu08}.
IC has been developed in many different language
settings~\citep{Shankar07, HammerAcRaGh07, Hammer09:ceal,
  Chen14:implicit}, and has addressed open problems in computational
geometry~\citep{AcarCoHuTu10}.

Some PL approaches to IC are \emph{static}, transforming programs to
derive a second program that can process input changes.
Static approaches perform these transformations a priori, before any
dynamic changes.  As such, static approaches are often lack the
ability to transform general recursion or to fully cache and exploit
dynamic dependencies~\citep{LiuTeitelbaum95, LiuStTe98, Cai2014}.

In contrast to static approaches, dynamic approaches attempt to trade
space for time savings.
A variety of dynamic approaches to IC have been proposed.
Most early approaches fall into one of two camps: they
either perform function caching of pure
programs~\citep{%
  Bellman57,McCarthy63,michie:1968,PughThesis%
}, or they
support input mutation and employ some form of dynamic dependency
graphs, along with a mechanism for performing \emph{change propagation}~\citep{%
  AcarBlHa04,DBLP:journals/entcs/AcarBBHT06,HammerAc08,AcarLW09,NominalAdapton2015%
}.
Earlier work restricted programs to those expressible as \emph{attribute
  grammars}~\citep{DemersReTe81,Reps82a,Reps82b,vogt1991}.
Various threads of research propose general schemes for practical
memoization, either making it applicable in more settings, or more
efficient.
Researchers have extended memoization to
parallel C and C{+}{+} programs~\citep{Bhatotia2015} and to distributed, cloud-based settings
\citep{Bhatotia2011}, and have reduced the (often large) space
overhead \citep{Chen2014}.

\section{Conclusion}

We have presented the complete implementation of microAdapton, a minimal system for incremental computation, and the higher-level miniAdapton interface built on top of it.
Like full Adapton, microAdapton and miniAdapton allow programmers to safely combine memoization and mutation.

Our approach of dividing our implementation into a ``mini''-level built on top a core ``micro''-level is inspired by microKanren and miniKanren.  As with microKanren, we take care to separate the hygienic macros in miniAdapton from the ``micro'' core.  This separation, and the careful exposition of the microKanren code in \citet{Hemann:SW:2013}, has resulted in readers of that paper porting microKanren to several dozen languages in addition to Scheme~\citep{minikanren_dot_org}.  We hope that microAdapton and miniAdapton will make the ideas and implementation of incremental computation similarly accessible, and will also result in readers porting the system to other languages and extending the system. One such extension would be to add cycle detection for the DCG.
\fixit{GitHub repo somewhere}

\section*{Acknowledgements}
This material is partially based on research sponsored by DARPA under
agreement number AFRL FA8750-15-2-0092 and by NSF under CAREER grant
1350344.  The views expressed are those of the authors and do not reflect the
official policy or position of the Department of Defense or the U.S.
Government. The U.S. Government is authorized to reproduce and
distribute reprints for Governmental purposes notwithstanding any
copyright notation thereon.

\fixit{Cite Dan and Jason proper}
We would also like to thank Jason Hemann and Dan Friedman for their work on the original microKanren system, which was a significant inspiration for this paper.

We also thank the Scheme workshop reviewers for their helpful comments.

\fixit{Acknowledge Scheme workshop reviewers (and especially Reviewer 7C for his time and attention) and organizers.}

\bibliography{paper}

\appendix
\fixit{There are still a few things to be fixed here}
\Matt{Are they fixed yet?}
\fixit{That was a very unhelpful comment on my part. I moved around the sections, and added a bit to the appendices. I'm having Will look at the appendices since so far we've been focusing our editing on the rest of the paper.}
\section{Complete microAdapton Implementation}
\label{sec:microAdaptonimplementation}
This appendix presents the full code for microAdapton.
microAdapton requires an appropriate implementation of sets---see Appendix~\ref{sec:sets} for a compatible set implementation using lists, and a brief performance discussion.
\begin{alltt}
(define-record-type
  (adapton adapton-cons adapton?)
  (fields
   thunk
   (mutable result)
   (mutable sub)
   (mutable super)
   (mutable clean?)))

(define (make-athunk thunk)
  (adapton-cons thunk
                'empty
                empty-set
                empty-set
                #f))

(define (adapton-add-dcg-edge! a-super a-sub)
  (adapton-sub-set! a-super
   (set-cons a-sub (adapton-sub a-super)))
  (adapton-super-set! a-sub
   (set-cons a-super (adapton-super a-sub))))

(define (adapton-del-dcg-edge! a-super a-sub)
  (adapton-sub-set! a-super
   (set-rem a-sub (adapton-sub a-super)))
  (adapton-super-set! a-sub
   (set-rem a-super (adapton-super a-sub))))

(define (adapton-compute a)
  (if (adapton-clean? a)
      (adapton-result a)
      (begin
        (set-for-each
          (lambda (x)
            (adapton-del-dcg-edge! a x))
          (adapton-sub a))
        (adapton-clean?-set! a #t)
        (adapton-result-set! a
          ((adapton-thunk a)))
        (adapton-compute a))))

(define (adapton-dirty! a)
  (when (adapton-clean? a)
        (adapton-clean?-set! a #f)
        (set-for-each adapton-dirty!
                      (adapton-super a))))

(define (adapton-ref val)
  (letrec ((a (adapton-cons
               (lambda () (adapton-result a))
               val
               empty-set
               empty-set
               #t)))
    a))

(define (adapton-ref-set! a val)
  (adapton-result-set! a val)
  (adapton-dirty! a))
\end{alltt}
\section{Complete miniAdapton Implementation}
\label{sec:miniAdaptonImplementation}
This appendix presents the full code for miniAdapton.
miniAdapton requires a memoization implementation---see Appendix~\ref{sec:memo} for a compatible memoization implementation using association lists, and a brief performance discussion.
\begin{alltt}
(define adapton-force
  (let ((currently-adapting #f))
    (lambda (a)
      (let ((prev-adapting
             currently-adapting))
        (set! currently-adapting a)
        (let ((result (adapton-compute a)))
          (set! currently-adapting
                prev-adapting)
          (when currently-adapting
                (adapton-add-dcg-edge!
                 currently-adapting
                 a))
          result)))))
\end{alltt}
\newpage
\begin{alltt}
(define-syntax adapt
  (syntax-rules ()
    ((_ expr)
     (make-athunk (lambda () expr)))))

(define (adapton-memoize-l f)
  (memoize (lambda x (adapt (apply f x)))))

(define (adapton-memoize f)
  (let ((f* (adapton-memoize-l f)))
    (lambda x (adapton-force (apply f* x)))))

(define-syntax lambda-amemo-l
  (syntax-rules ()
    ((_ (args ...) body ...)
     (let ((f* (adapton-memoize-l
                (lambda (args ...)
                  body ...))))
       (lambda (args ...) (f* args ...))))))

(define-syntax lambda-amemo
  (syntax-rules ()
    ((_ (args ...) body ...)
     (let ((f* (adapton-memoize
                (lambda (args ...)
                  body ...))))
       (lambda (args ...) (f* args ...))))))

(define-syntax define-amemo-l
  (syntax-rules ()
    ((_ (f args ...) body ...)
     (define f (lambda-amemo-l (args ...)
                 body ...)))))

(define-syntax define-amemo
  (syntax-rules ()
    ((_ (f args ...) body ...)
     (define f (lambda-amemo (args ...)
                 body ...)))))

(define-syntax define-avar
  (syntax-rules ()
    ((_ name expr)
     (define name
             (adapton-ref (adapt expr))))))

(define (avar-get v)
  (adapton-force (adapton-force v)))

(define-syntax avar-set!
  (syntax-rules ()
    ((_ v expr)
     (adapton-ref-set! v (adapt expr)))))
\end{alltt}
\section{Spreadsheet Example Redux}
The motivating example for the original Adapton
work \citep{Hammer:2014:ACD:PLDI} consists of a incremental
``spreadsheet'' evaluator, given as a simple interpreter, made
incremental via use of the Adapton primitives.
In this section, we show that miniAdapton can also express this
example.

In the original example, given in a variant of ML, the programmer
specifies a data structure to represent expressions that change over
time, and a recursive function that evaluates these expressions to
their valuation.
In this version, we do not employ an explicit data structure
representation for expressions; rather, we use the host Scheme
evaluator in place of this programmer-defined interpreter.
This approach employs avars to hold (Scheme) expressions that can
change over time; in particular, avars hold suspended computations
that represent the formulae of the spreadsheet cells.
In this way, each avar is like a cell in a (very) simple spreadsheet.
Under this representation, we can encode the example from the prior
work as follows:

\begin{alltt}
(define-avar n1 1)
(define-avar n2 2)
(define-avar n3 3)
(define-avar p1 (+ (avar-get n1) (avar-get n2)))
(define-avar p2 (+ (avar-get p1) (avar-get n3)))
\end{alltt}

This code creates five cells in the spreadsheet, which hold three
constants (\texttt{1}, \texttt{2} and \texttt{3}) and two sums over
these ({\tt p1} and {\tt p2}).

With these bindings in place, we can encode the example session as
follows:
\begin{alltt}
> (avar-get p1)
3
> (avar-get p2)
6
\end{alltt}

Because Adapton (and miniAdapton) are demand-driven, the evaluation
of \texttt{p1} and \texttt{p2} are suspended until they are forced to
execute, by the uses of \texttt{avar-get} above; these operations
compute their valuations, 3 and 6, respectively.

\begin{alltt}
> (avar-set! n1 5)
> (avar-get p1)
7
\end{alltt}

Next, the user mutates an avar using \texttt{avar-set!}, and
re-observes the valuation of \texttt{avar-get}.
Meanwhile, the valuation of \texttt{n2} is unaffected, and need not be
recomputed.  In this case, \texttt{n2} is merely a constant value, but
if it were more complex, this entire computation would still be
reusable.
This demonstrates how miniAdapton allows memoization and mutation to
safely coexist.

As with Adapton, if the user decides to mutate the input
by \emph{swapping} expression trees, miniAdapton responds by swapping
their corresponding memoized sub-computations behind the scenes:

\begin{alltt}
> (avar-set! p2 (+ (avar-get n3) (avar-get p1)))
> (avar-get p2)
10
\end{alltt}

Finally, the following user interactions demonstrate how miniAdapton
permits memoization to benefit from \emph{switching}, where the user
updates avar \texttt{p1}, but then changes their mind and reverts its
expression:

\begin{alltt}
> (avar-set! p1 4)
> (avar-get p2)
7
> (avar-set! p1 (+ (avar-get n1) (avar-get n2)))
> (avar-get p2)
10
\end{alltt}

In this case, the original computation (with \texttt{p1}
holding \mbox{\texttt{(+ (avar-get n1) (avar-get n2))}}) is recovered from
the memoized cache.

By using Scheme's evaluator, rather than a special
language that we define (as in the original Adapton work), our
spreadsheet language can use primitives from Scheme, rather than
having to encode them.
In effect, we acquire the spreadsheet formulae
language from Scheme, merely by using avars.
As a simple example, we can immediately use other
primitives (like multiplication) without having to redefine their
meaning for this spreadsheet language:

\begin{alltt}
> (avar-set! p1 (* (avar-get n1) (avar-get n2)))
10
> (avar-get p2)
13
\end{alltt}

Doing the equivalent in the original paper would have required
modifying the interpreter, but here, Scheme's standard multiplication
function suffices.

\section{Set Implementation}
\label{sec:sets}
The microAdapton implementation in Appendix~\ref{sec:microAdaptonimplementation} requires an implementation of sets.
This appendix presents a minimal but sufficient list-based set implementation which yields the desired semantics.
Most operations take linear time on average, which can cause microAdapton to run slowly when adding or deleting edges in an athunk that has large numbers of ingoing or outgoing nodes.
More efficient set representations include self-balancing trees and hashtables.
Mutable set implementations are also compatible with the microAdapton implementation in Appendix~\ref{sec:microAdaptonimplementation}, provided the microAdapton code is modified for the mutable interface.
\begin{alltt}
(define empty-set '())
(define (set-mem e s)
  (memv e s))
(define (set-cons e s)
  (if (set-mem e s) s (cons e s)))
(define (set-rem e s)
  (filter (lambda (x) (not (eqv? e x))) s))
(define (set-union s1 s2)
  (fold-left set-cons s2 s1))
(define (set-intersect s1 s2)
  (fold-left set-rem s2 s1))
(define set-for-each for-each)
(define set->list (lambda (x) x))
\end{alltt}
\section{Memoization}
\label{sec:memo}
This appendix presents an implementation of memoization, compatible with the miniAdapton implementation in Appendix~\ref{sec:miniAdaptonImplementation}.
\begin{alltt}
(define (memoize f)
  (let ((s (make-kv-store)))
    (lambda x ; variadic memoization
      (let ((k/v (lookup-kv-store s x)))
        (if k/v
            (cdr k/v)
            (let ((result (apply f x)))
              (add-kv-store! s x result)
              result))))))

(define simple-memoization-test
  (memoize (lambda x (read))))

(define-syntax lambda-memo
  (syntax-rules ()
    ((_ (args ...) body ...)
     (let ((f* (memoize
                (lambda (args ...)
                  body ...))))
       (lambda (args ...) (f* args ...))))))

(define-syntax define-memo
  (syntax-rules ()
    ((_ (f args ...) body ...)
     (define f (lambda-memo (args ...)
                 body ...)))))
\end{alltt}

The memoization code requires a key-value store implementation, such as this one, which uses association lists:
\begin{alltt}
(define (make-kv-store) (list '()))
(define (add-kv-store! s k v)
  (set-car! s `((,k . ,v) . ,(car s))))
(define (lookup-kv-store s k)
  (assoc k (car s)))
\end{alltt}
This key-value store implementation is inefficient, taking linear time looking up any item not currently in the store.
For reasonable performance a more efficient key-value store implementation should be used.
Most memoization implementations use hash tables, which typically take constant time for most operations.
In addition, our implementation uses \texttt{equal?} (through \texttt{assoc}) where the more efficient pointer equality tests \texttt{eq?} or \texttt{eqv?} might be preferred.

\end{document}